# BEAM PERFORMANCE AND LUMINOSITY LIMITATIONS IN THE HIGH-ENERGY STORAGE RING (HESR)

A. Lehrach[*], O. Boine-Frankenheim[#], F. Hinterberger[+], R. Maier[*], and D. Prasuhn[*]

[*]*Forschungszentrum Jülich (FZJ), Jülich, Germany*
[#]*Gesellschaft für Schwerionenforschung (GSI), Darmstadt, Germany*
[+]*Helmholtz Institut für Strahlen- und Kernphysik (HISKP), Universität Bonn, Germany*

*Abstract*

The High-Energy Storage Ring (HESR) of the future International Facility for Antiproton and Ion Research (FAIR) at GSI in Darmstadt is planned as an antiproton synchrotron storage ring in the momentum range of 1.5 to 15 GeV/c. An important feature of this new facility is the combination of phase space cooled beams and dense internal targets (e.g. pellet targets), which results in demanding beam parameter requirements for two operation modes: high luminosity mode with peak luminosities of up to $2 \cdot 10^{32}$ cm$^{-2}$ s$^{-1}$, and high resolution mode with a momentum spread down to $10^{-5}$, respectively. To reach these beam parameters one needs a very powerful phase space cooling, utilizing high-energy electron cooling and high-bandwidth stochastic cooling. The effects of beam-target scattering and intra-beam interaction are investigated in order to study beam equilibria and beam losses for the two different operation modes.

## INTRODUCTION

The HESR is dedicated to the field of high-energy antiproton physics, to explore the research areas of charmonium spectroscopy, hadronic structure, and quark-gluon dynamics with high-quality beams over a broad momentum range from 1.5 to 15 GeV/c [1,2]. According to the Conceptual Design Report (CDR) [1] the HESR was planned with only one internal interaction point, equipped with the PANDA detector [3]. Two other experimental groups (ASSIA [4] and PAX [5,6]) also expressed interest in spin physics experiments at the HESR. This requires a synchrotron mode to accelerate polarized beams in the HESR.

## DESIGN ISSUES AND EXPERIMENTAL REQUIREMENTS

The HESR lattice is designed as a racetrack-shaped storage ring, consisting of two 180° arc sections connected by two long straight sections (see Fig. 1). One straight section will mainly be occupied by the electron cooler. In a later stage a Siberian snake can be installed to preserve polarization during acceleration [7]. The other straight section will host the experimental installation with internal frozen $H_2$ pellet jet target, injection kickers/septa and RF cavities. Two pickup tanks for stochastic cooling are located close to the ends of one straight section while the stochastic kicker tanks are placed opposite in the other straight section, diagonally connected with signal lines. Special requirements for the lattice are dispersion-free straight sections and small betatron amplitude of about 1 m at the internal interaction point, imaginary transition energy, and optimized ion optical conditions for beam cooling (e.g. matched betatron amplitudes at the pickups and kickers of the stochastic cooling system and in the electron cooler section). Details of the ion optical layout and features of the lattice design are discussed in [8]. The antiproton beam is accumulated in the CR/RESR complex at 3.8 GeV/c [9]. Beam parameters depend on the number of accumulated particles.

TABLE 1. Beam parameters and operation modes.

| Injection Parameters | |
|---|---|
| Transverse emittance | 1 mm·mrad (normalized, rms) for $3.5 \cdot 10^{10}$ particles, scaling with number of accumulated particles: $\varepsilon_\perp \sim N^{4/5}$ |
| Relative momentum spread | $1 \cdot 10^{-3}$ (normalized, rms) for $3.5 \cdot 10^{10}$ particles, scaling with number of accumulated particles: $\sigma_p/p \sim N^{2/5}$ |
| Bunch length | 200 m |
| Injection Momentum | 3.8 GeV/c |
| Injection | Kicker injection using multi-harmonic RF cavities |
| **Experimental Requirements** | |
| Ion species | Antiprotons |
| Production rate | $2 \cdot 10^7$ /s ($1.2 \cdot 10^{10}$ per 10 min) |
| Momentum / Kinetic energy range | 1.5 to 15 GeV/c / 0.83 to 14.1 GeV |
| Number of particles | $10^{10}$ to $10^{11}$ |
| Target thickness | $4 \cdot 10^{15}$ atoms/cm$^2$ |
| Transverse emittance | 1 to 2 mm·mrad |
| Betatron amplitude at interaction point | 1 m |
| **Operation Modes** | |
| High resolution (HR) | Luminosity of $2 \cdot 10^{31}$ cm$^{-2}$ s$^{-1}$ for $10^{10}$ $\bar{p}$ rms momentum spread $\sigma_p/p \sim 10^{-5}$ 1.5 to 9 GeV/c, electron cooling |
| High luminosity (HL) | Luminosity of $2 \cdot 10^{32}$ cm$^{-2}$ s$^{-1}$ for $10^{11}$ $\bar{p}$ rms momentum spread $\sigma_p/p \sim 10^{-4}$ 1.5 to 15 GeV/c, stochastic cooling above 3.8 GeV/c |

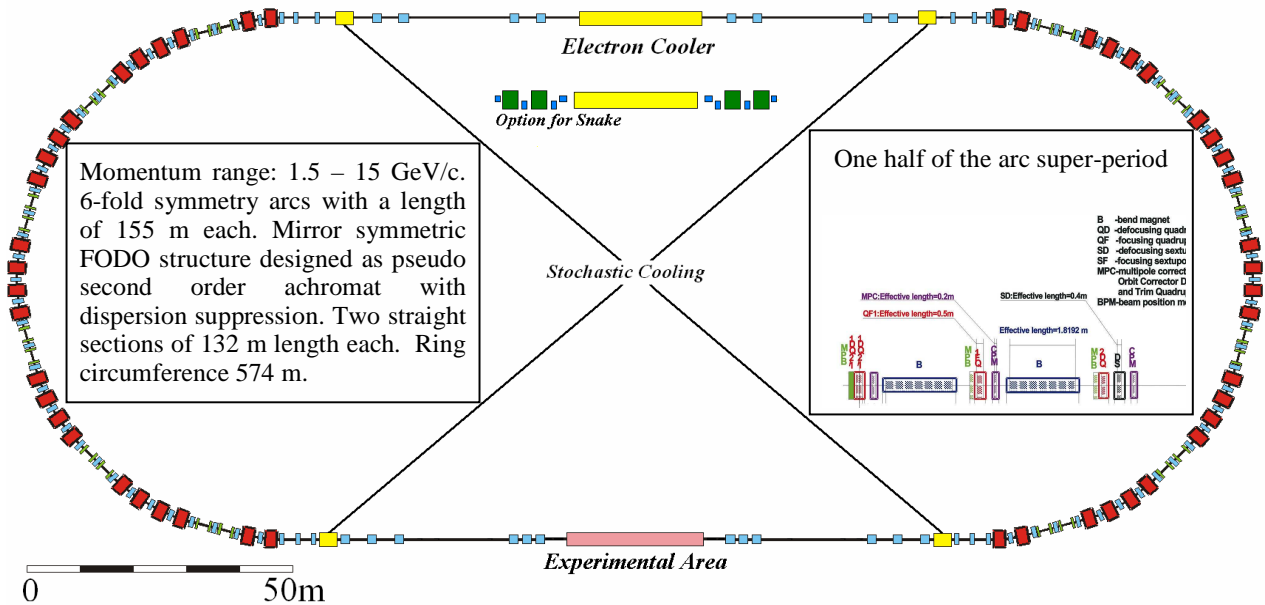

FIGURE 1. Schematic view of the HESR with 6-fold symmetry lattice. Tentative positions for injection, cooling devices and experimental installations are indicated. Also shown is the arrangement of elements in the super-period.

Table 1 summarizes the specified injection parameters, experimental requirements and operation modes. Demanding requirements for high intensity and high quality beams are combined in two operation modes: high luminosity (HL) and high resolution (HR), respectively. The high-resolution mode is defined in the momentum range from 1.5 to 9 GeV/c. To reach a momentum resolution down to $\sigma_p/p \sim 10^{-5}$, only $10^{10}$ circulating particles in the ring are anticipated. The high-luminosity mode requires an order of magnitude higher beam intensity with reduced momentum resolution to reach a peak luminosity of $2 \cdot 10^{32}$ cm$^{-2}$ s$^{-1}$ in the full momentum range.

## BEAM COOLING SYSTEMS AND TARGETS

A feasibility study for magnetized high-energy electron cooling was presented by the Budker Institute for Nuclear Physics (BINP) [10]. An electron beam up to 1 A, accelerated in special accelerator columns to energies in the range of 0.45 to 8 MeV, is proposed for the HESR. The 30 m long solenoidal field in the cooler section ranges from 0.2 to 0.5 T with a magnetic field straightness on the order of $10^{-5}$. Recently it was decided, that electron cooling should only cover the momentum range of the high-resolution mode, leading to maximum beam energy of 4.5 MeV. Further design work on the electron cooler is lead by The Svedberg Laboratory (TSL) in Uppsala in cooperation with other institutes including the Budker Institute, Fermi National Accelerator Laboratory (FNAL) and industry [11]. The main stochastic cooling parameters were determined for a system utilizing quarter-wave loop pickups and kickers [2]. Stochastic cooling is presently specified above 3.8 GeV/c. This beam cooling method has the advantage of being capable to separately cool the transverse and longitudinal phase space.

Frozen $H_2$ pellets are required to reach the specified target thickness of $4 \cdot 10^{15}$ atoms/cm$^2$. The pellet size is 20 to 40 μm in diameter. The pellets velocity is 60 m/s with a flow rate of 20000 pellets/s, leading to an average longitudinal distance between pellets of a few millimeters. The pellet stream moves with an angular divergence of ±0.04° corresponding to a transverse position uncertainty of ±1 mm at the interaction point [12]. For a betatron amplitude of 1 m, a beam emittance on the order of 1 mm·mrad is required to ensure sufficient beam-target overlap.

## COOLED BEAM EQUILIBRIA

Beam equilibrium is of a major concern for the high-resolution mode. Calculations of beam equilibrium between electron cooling, intra-beam scattering and beam-target interaction are being performed utilizing different simulation codes like BETACOOL by I.N. Meshkov et al. (JINR, Dubna), MOCAC by A.E. Bolshakov et al. (ITEP, Moscow), and PTARGET by B. Franzke at al. (GSI, Darmstadt). Results from different codes for HESR conditions are compared in [13]. Studies of beam equilibria for the HESR are also carried out by D. Reistad for electron cooled beams [2] and by H. Stockhorst for stochastically cooled beams [14] utilizing the BETACOOL code.

To simulate the dynamics of the core particles, an analytic rms model was applied for the calculation presented in this paper [15]. The empirical magnetized cooling force formula by V.V Parkhomchuk was used for electron cooling [16], and an analytical description was applied for intra-beam scattering [17]. Beam heating by beam-target interaction is described by transverse and longitudinal emittance growth due to Coulomb scattering and energy straggling, respectively [18,19]. Electron cooler and target parameters for these simulations are summarized in table 2.

TABLE 2. Electron cooler and target parameters.

| Electron Cooler | |
|---|---|
| Length of cooling section | 30 m |
| Electron current | 0.2 A |
| Effective velocity | $2 \cdot 10^4$ m/s |
| Betatron amplitude at Cooler | 100 m |
| Pellet Target | |
| Target density | $4 \cdot 10^{15}$ atoms/cm$^2$ |
| Betatron amplitude at target | 1 m |

Transverse emittance and momentum spread in equilibrium are plotted versus beam energy for a luminosity of $2 \cdot 10^{31}$ cm$^{-2}$ s$^{-1}$ in Fig. 2.

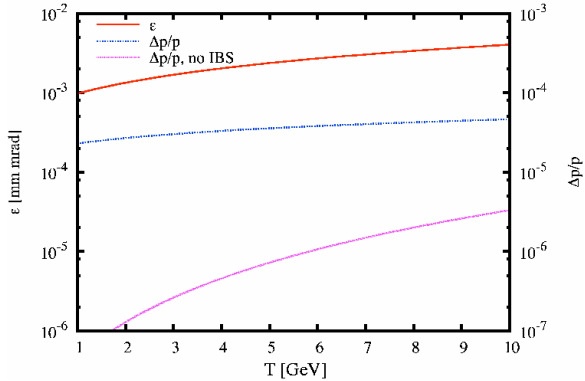

FIGURE 2. Transverse rms beam emittance (upper curve) and momentum spread (two lower curves) in equilibrium vs. kinetic beam energy $T$ for a luminosity of $2 \cdot 10^{31}$ cm$^{-2}$ s$^{-1}$. Momentum spread equilibria with and without intra-beam scattering are plotted.

Transverse rms beam emittances of about $10^{-3}$ up to a few times $10^{-2}$ mm·mrad and rms momentum spreads as low as $3 \cdot 10^{-5}$ can be reached in the energy range of the high-resolution mode. The calculations show that the beam equilibria are dominated by intra-beam scattering. Beam heating by the target is at least one order of magnitude weaker. The high-resolution mode seems not feasible under these conditions since the equilibrium momentum spread is larger than specified. Equilibrium beam emittances also do not provide a sufficient beam-target overlap, external transverse beam heating is required.

To study the dynamics of equilibrium momentum spread in case of larger transverse beam size, the transverse beam emittance was artificially kept constant at the level of $10^{-1}$ mm·mrad in the simulation (see Fig. 3). Momentum spreads down to $10^{-5}$ seem feasible in this case. For a sufficient beam-target overlap, the transverse beam emittance has even to be one order of magnitude larger. Further studies have to be carried out to reach beam emittances on the order of 1 mm·mrad.

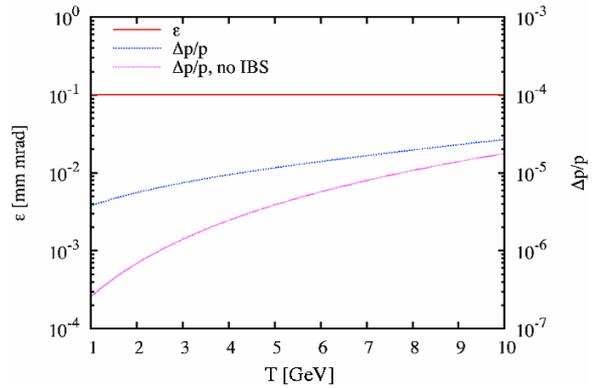

FIGURE 3. Momentum spread (two lower curves) in equilibrium vs. kinetic beam energy $T$ for a luminosity of $2 \cdot 10^{31}$ cm$^{-2}$ s$^{-1}$. Momentum spread equilibria with and without intra-beam scattering are plotted. The transverse beam emittance is artificially kept constant (upper curve).

The presently available numerical models to calculate beam equilibrium parameters of cooled beams interacting with internal targets will be further improved. Important tasks covered by an INTAS research project are studies of beam equilibria, and cooled beam distributions in the presence of intra-beam scattering and internal target scattering [20].

The beam intensity and lifetime of the circulation beam can also be limited by the electron beam, due to its defocusing effect on the circulation antiproton beam, and coherent instabilities caused by positive residual gas ions trapped in the potential of the electron beam [21,22,23].

## BEAM LOSSES

The main restriction for high luminosities is beam losses, since the antiproton production rate is limited. Three dominating contributions of the beam-target interaction have been identified: Hadronic interaction, single Coulomb scattering and energy straggling of the circulating beam in the target. In addition, single intra-beam scattering due to the Touschek effect has also to be considered for beam lifetime estimations. Beam losses due to residual gas scattering can be neglected compared to beam-target interaction, if the vacuum is better than $10^{-9}$ mbar.

The relative beam loss rate for the total cross section $\sigma_{tot}$ is given by the expression

$$(\tau_{loss}^{-1}) = n_t \sigma_{tot} f_0, \qquad (1)$$

where $\tau_{loss}$ is the $1/e$ beam lifetime, $n_t$ the target thickness and $f_0$ the reference particle's revolution frequency.

*Hadronic interaction*

The total hadronic cross section is shown in Fig. 4 versus beam momentum. The total cross section decreases roughly from 100 mbarn at 1.5 GeV/c, to 57 mbarn at 9 GeV/c, and to 51 mbarn at 15 GeV/c. Based on this numbers and revolution frequencies of 443, 519 and 521 kHz, relative beam loss rates are estimated to be $1.8 \cdot 10^{-4}$ /s at 1.5 GeV/c, $1.2 \cdot 10^{-4}$ /s at 9 GeV/c, and $1.1 \cdot 10^{-4}$ /s at 15 GeV/c for a target thickness of $4 \cdot 10^{15}$ atoms/cm$^2$.

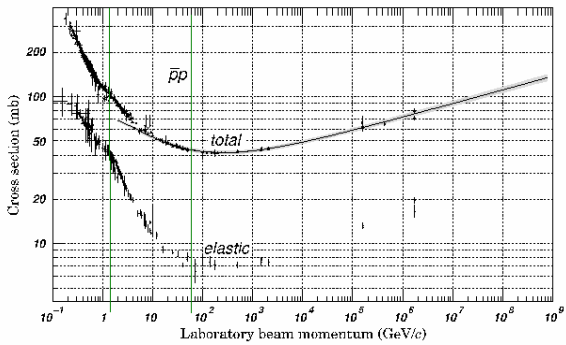

FIGURE 4. Total and elastic cross section for $p - \bar{p}$ collisions as a function of laboratory beam momentum (from http://pdg.lbl.gov/xsect/contents.html).

*Single Coulomb scattering*

Coulomb scattering is described by the Rutherford cross section. Small angle scattering can be compensated by phase space cooling. Particles single scattered out of the transverse acceptance are lost. The cross section for single Coulomb scattering is given by

$$\sigma_{tot} = \frac{4\pi Z_t^2 Z_i^2 r_i^2}{\beta_0^4 \gamma_0^2 \theta_{acc}^2}, \qquad (2)$$

where $Z_t$ and $Z_i$ are the charge numbers of target and projectile, $r_i = 1.535 \cdot 10^{-16}$ cm is the classical proton radius, $\beta_0$ and $\gamma_0$ are the kinematic parameters of the circulating beam. For angles larger than the acceptance angle $\theta_{acc}$ scattered particles are lost

$$\theta_{acc} = \sqrt{\frac{A}{\beta}}. \qquad (3)$$

The transverse acceptance $A$ is related to the beam emittance providing sufficient beam-target overlap. $\beta$ is the betatron amplitude at the interaction point. Without beam cooling, one could assume that scattered particles with a transverse emittance larger than 1 mm·mrad do not contribute to the luminosity any more and can be treated as lost. The relative loss rate would then range from $2.9 \cdot 10^{-4}$ /s at 1.5 GeV/c, $6.8 \cdot 10^{-6}$ at 9 GeV/c to $2.4 \cdot 10^{-6}$ /s at 15 GeV/c. This is the upper limit for beam losses. Single Coulomb scattering would be the dominant mechanism for low momenta. Due to beam cooling, scattered particles can be cooled back onto the target. For an electron beam with a diameter of 10 mm in the cooling section and a betatron amplitude of 100 m at the electron cooler, particles scattered from the target would also be affected by the non-linear field of the electron beam and would most probably be lost. Detailed simulations are needed to find optimum parameters for target and electron beam diameter to reach larger values of maximum beam emittances still contributing to the luminosity. Stochastic cooling could certainly be a very powerful tool to suppress beam losses due to Coulomb scattering if it is made available at lower momentum of the HESR, maybe in combination with electron cooling.

*Energy loss straggling*

Energy loss due to beam-target interaction out of the longitudinal acceptance of the accelerator leads to beam losses. The single collision-energy loss probability (with the energy loss $\varepsilon$) can be described by a Rutherford-like expression

$$w(\varepsilon) = \frac{\xi}{\varepsilon^2}\left(1 - \beta_0^2 \frac{\varepsilon}{\varepsilon_{max}}\right), \qquad (4)$$

with a maximum energy transfer of

$$\varepsilon_{max} = \frac{2 m_e c^2 \beta_0^2 \gamma_0^2}{1 + 2\gamma_0 \dfrac{m_e}{m_i} + \left(\dfrac{m_e}{m_i}\right)^2}, \qquad (5)$$

the electron mass $m_e$ and incident particle (antiproton) mass $m_i$ [24]. The scaling factor reads

$$\xi = 153.4 \frac{\text{keV}}{\text{g/cm}^2} \frac{Z_i^2}{\beta_0^2} \frac{Z_t}{A_t} \rho x. \qquad (6)$$

Here, $A_t$ is the mass number of the target and $\rho x$ the target density times the effective target thickness. The second moment of the energy loss probability yields the mean square energy deviation $\Delta\varepsilon_{rms}^2$. The corresponding mean square relative momentum deviation is given by

$$\delta_{rms}^2 = \left(\frac{\gamma_0}{\gamma_0+1}\right)^2 \frac{\Delta\varepsilon_{rms}^2}{T_0^2}, \quad (7)$$

where $T_0$ is the kinetic energy of the reference particle. By integrating over the probability function one gets the relative beam loss rate

$$(\tau_{loss,\parallel}^{-1})_S = f_0 \int_{\varepsilon_{cut}}^{\varepsilon_{max}} w(\varepsilon)d\varepsilon$$
$$= f_0 \xi \left(\frac{1}{\varepsilon_{cut}} - \frac{1}{\varepsilon_{max}} - \frac{\beta_0^2}{\varepsilon_{max}} \ln\frac{\varepsilon_{max}}{\varepsilon_{cut}}\right). \quad (8)$$

Assuming $\delta_{cut} = \left(\frac{\gamma_0}{\gamma_0+1}\right)\frac{\Delta\varepsilon_{cut}}{T_0} = 10^{-3}$, the relative beam loss probability per turn is shown in Fig. 5. Relative loss rates are ranging from $1.3\cdot10^{-4}$ /s at 1.5 GeV/c, $4.1\cdot10^{-5}$ at 9 GeV/c to $2.8\cdot10^{-5}$ /s at 15 GeV/c.

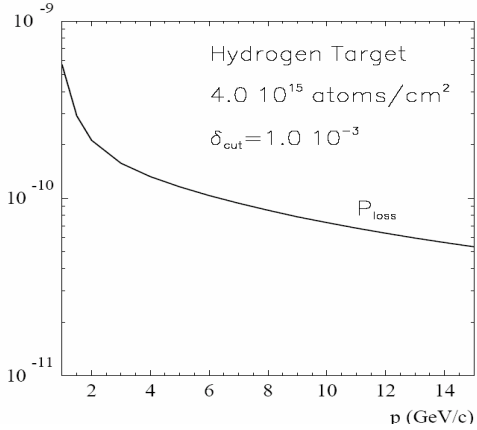

FIGURE 5. Loss probability per turn $\int w(\varepsilon)d\varepsilon$ vs. beam momentum assuming $\delta_{cut} = 10^{-3}$.

## Touschek effect

For small transverse emittances, the beam can be lost due to single large-angle intra-beam scattering in the longitudinal ring acceptance [17]. The beam loss rate is determined by the longitudinal diffusion coefficient

$$D_\parallel^{IBS} = \frac{\Lambda_\parallel^{IBS}}{\varepsilon_\perp^{3/2}}, \quad \Lambda_\parallel^{IBS} = \frac{\sqrt{\pi}N_i c r_i^2 L_c}{4\gamma_0^3 \beta_0^3 \langle\beta_\perp^{1/2}\rangle C}. \quad (9)$$

$\varepsilon_\perp$ is the transverse rms beam emittance, $N_i$ is the number of circulating ions, $c$ is the speed of light, $L_c \approx 10$ is the Coulomb logarithm, $\langle\beta_\perp^{1/2}\rangle = \sqrt{7.5\ m}$ is the average of the square root of the betatron amplitude in the ring and $C$ is the ring's circumference. The relative beam loss rate then reads

$$(\tau_{loss}^{-1})_{IBS} = \frac{D_\parallel^{IBS}}{L_C \delta_{cut}^2}, \quad (10)$$

where $\delta_{cut}$ is the longitudinal ring acceptance. In table 3 the relative beam loss rate is listed for different beam momenta and transverse beam emittances assuming a longitudinal acceptance of $\delta_{cut} = 10^{-3}$.

TABLE 3. Relative beam loss rate due the Touschek effect assuming $N_i = 10^{11}$ for different emittances.

| | $(\tau_{loss}^{-1})$ / s$^{-1}$ | | |
|---|---|---|---|
| rms emittance | 1.5 GeV/c | 9 GeV/c | 15 GeV/c |
| 0.01 mm·mrad | 4.9·10$^{-2}$ | 2.3·10$^{-4}$ | 4.9·10$^{-5}$ |
| 1 mm·mrad | 4.9·10$^{-5}$ | 2.3·10$^{-7}$ | 4.9·10$^{-8}$ |

As expected, the beam loss rate is rather large for small transverse emittance and decreases with beam momentum by the third power. For the equilibrium emittance this effect would be dominating at low momentum. Since a larger emittance is needed for sufficient beam-target overlap anyhow, this loss rate will be small compared to other effects.

## Beam lifetime

For beam-target interaction, the beam lifetime is independent of the beam intensity, whereas for the Touschek effect it depends on the beam equilibria and therefore on the beam intensity. The total relative loss rate is given by

$$(\tau_{loss}^{-1}) = (\tau_{loss}^{-1})_H + (\tau_{loss}^{-1})_C + (\tau_{loss}^{-1})_S + (\tau_{loss}^{-1})_{IBS}. \quad (11)$$

In table 4 the upper limit for beam losses and corresponding lifetimes are listed for a transverse beam emittance of 1 mm·mrad and $10^{11}$ circulating particles.

TABLE 4. Upper limit for relative beam loss rate and beam lifetime at different beam momenta.

| | $(\tau_{loss}^{-1})$ / s$^{-1}$ | | |
|---|---|---|---|
| Heating Process | 1.5 GeV/c | 9 GeV/c | 15 GeV/c |
| Hadronic Interaction | 1.8·10$^{-4}$ | 1.2·10$^{-4}$ | 1.1·10$^{-4}$ |
| Single Coulomb | 2.9·10$^{-4}$ | 6.8·10$^{-6}$ | 2.4·10$^{-6}$ |
| Energy Straggling | 1.3·10$^{-4}$ | 4.1·10$^{-5}$ | 2.8·10$^{-5}$ |

| Touschek Effect | $4.9 \cdot 10^{-5}$ | $2.3 \cdot 10^{-7}$ | $4.9 \cdot 10^{-8}$ |
|---|---|---|---|
| Total relative loss rate | $6.5 \cdot 10^{-4}$ | $1.7 \cdot 10^{-4}$ | $1.4 \cdot 10^{-4}$ |
| 1/e beam lifetime $t_{pbar}$ | ~ 1540 s | ~ 6000 s | ~ 7100 s |

For the discussed values, beam lifetimes are ranging from 1540 s to 7100 s for the high-luminosity mode. Less than half an hour beam lifetime is too small compared to the antiproton production rate. Beam lifetimes at low momenta strongly depend on the beam cooling scenario and ring acceptance. The beam loss rate for single Coulomb scattering could significantly be reduced by applying a larger electron beam diameter in combination with stochastic cooling. Beam lifetime in this case would increase to about 2000 s at 1.5 GeV/c. Also the longitudinal acceptance $\delta_{cut}$ should be at least a factor of two larger than $10^{-3}$ to reduce the effect of energy straggling and finally reach a beam lifetime close to one hour at 1.5 GeV/c.

Since beam losses due to single Coulomb scattering and energy straggling decrease very fast with beam momentum, already at 2.5 GeV/c the beam lifetime is close to an hour ($\tau = 3470\ s$). At higher beam momenta, beam losses are dominated by the hadronic interaction.

## AVERAGE LUMINOSITY

To calculate the average luminosity, machine cycles and beam preparation times have to be specified. After injection, the beam is pre-cooled to equilibrium (with target off). Pre-cooling at 3.8 GeV/c takes about 30 s to 60 s, depending on the initial beam parameters and beam cooling method. The beam is then ac-/decelerated to the desired beam momentum. A maximum ramp rate for the superconducting dipole magnets of 25 mT/s is specified, leading to acceleration duration of 100 s for maximum momentum. Beam steering and focusing in the target region takes about 10 s. Beam cooling and pellet beam are switched on before the physics experiment can be performed. A typical evolution of the luminosity during a cycle is plotted in Fig. 6 versus time in the cycle.

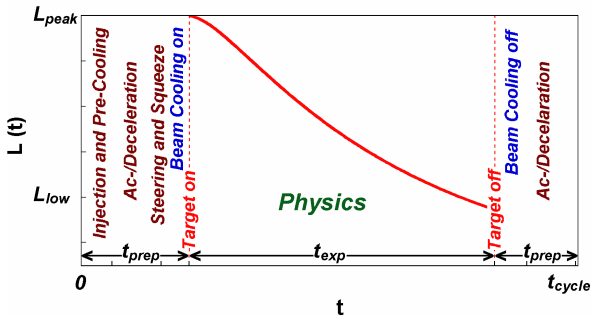

FIGURE 6. Time dependent luminosity during the cycle $L(t)$ versus time in cycle. Different measures for beam preparation are indicated.

Total beam preparation time $t_{prep}$ ranges from 120 s for 1.5 GeV/c to 290 s for 15 GeV/c. In the high-luminosity mode, particles should be re-used in the next cycle. Therefore the used beam is transferred back to the injection momentum and merged with the newly injected beam. A barrier bucket scheme is foreseen for the injection and beam accumulation procedure. During acceleration 1% and during deceleration 5% beam losses are assumed. The initial luminosity after beam preparation is given by

$$L_0 = f_0 N_{i,0} n_t, \qquad (12)$$

where $N_{i,0}$ is the number of available particles after the target is switched on. $N_{i,0}$ depends on the production rate, beam lifetime and beam preparation time. It is plotted versus different cycle times for in Fig. 7.

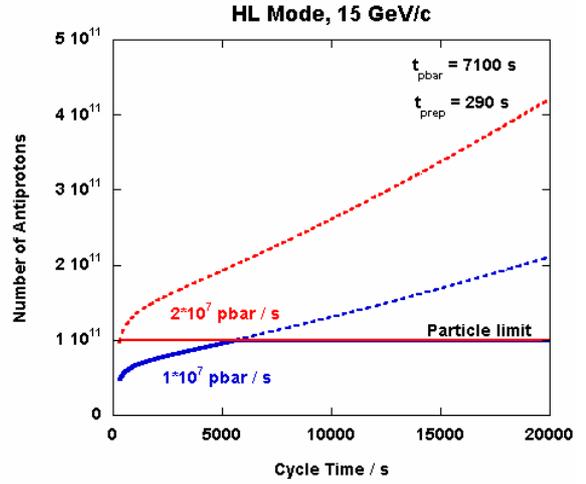

FIGURE 7. Number of initially available antiprotons $N_{i,0}$ for different cycle times at 15 GeV/c. The dashed lines show the number of antiprotons without any intensity restriction, the solid lines with the restriction of $N_{i,0} = 10^{11}$.

For a production rate of $2 \cdot 10^7$ antiprotons per second and a beam lifetime of 7100 s, $10^{11}$ particles can always be provided to the experiment. For a lower production rate of $1 \cdot 10^7$ antiprotons per second, the cycle time has to be longer than 5000 s to reach the required number of particles for the high-luminosity mode at 15 GeV/c. To calculate the average luminosity, one has to integrate the time dependent luminosity over the experimental time (beam on target) $t_{exp}$

$$\overline{L} = \frac{\int_0^{t_{exp}} L_0 e^{-\frac{t}{\tau}} dt}{t_{cycle}} = \frac{L_0 \tau \left[1 - e^{-\frac{t_{exp}}{\tau}}\right]}{t_{prep} + t_{exp}}, \qquad (13)$$

where $\tau$ is the beam lifetime, and $t_{cycle}$ the total time of the cycle, with $t_{cycle} = t_{exp} + t_{prep}$. The average luminosity can then be written as

$$\overline{L} = f_0 N_{i,0} n_t \frac{\tau[1 - e^{-\frac{t_{exp}}{\tau}}]}{t_{exp} + t_{prep}} \quad . \quad (14)$$

The dependence of the average luminosity on the cycle time is shown for different antiproton production rates in Figs. 8 and 9 for a given mean beam lifetime $t_{pbar}$ and preparation time $t_{prep}$.

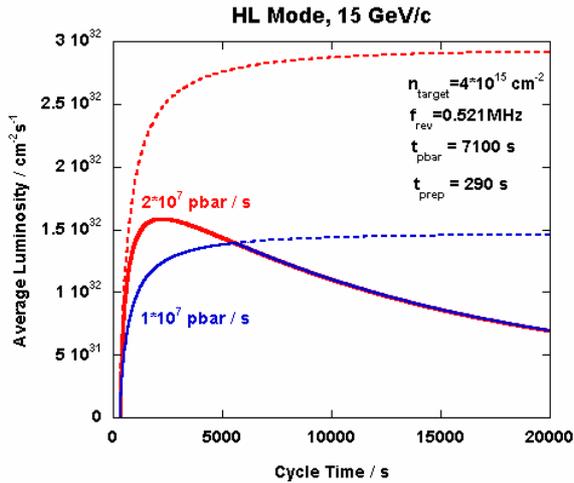

FIGURE 8. Average luminosity vs. cycle time at 15 GeV/c. The maximum number of particles is limited to $10^{11}$ for the solid line, and unlimited for the dashed lines.

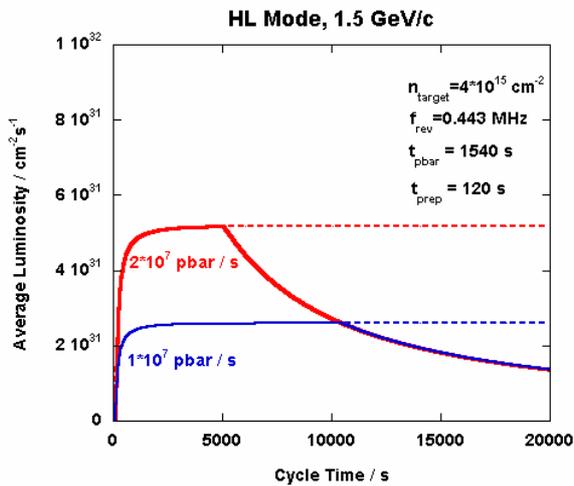

FIGURE 9. Average luminosity vs. cycle time at 1.5 GeV/c. The maximum number of particles is limited to $10^{11}$ for the solid line, and unlimited for the dashed lines.

With limited number of antiprotons ($N_{i,0} = 10^{11}$), as specified for the high-luminosity mode, average luminosities of up to $1.6 \cdot 10^{32}$ cm$^{-2}$ s$^{-1}$ are achieved at 15 GeV/c for cycle times of less than the beam lifetime. If one does not restrict the number of available particles, cycle times should be longer to reach maximum average luminosities close to $3 \cdot 10^{32}$ cm$^{-2}$ s$^{-1}$. This is a theoretical upper limit, since the momentum spread of the injected beam would be much bigger, leading to larger beam losses during injection due to the limited longitudinal ring acceptance. Also pre-cooling after injection would take longer and the specified momentum resolution for the high-luminosity mode of $\sigma_p/p \sim 10^{-4}$ would not be reached anymore. For the lowest momentum, more than $10^{11}$ particles can not be provided, due to very short beam lifetimes. Average luminosities are below $10^{32}$ cm$^{-2}$ s$^{-1}$ at 1.5 GeV/c. Already at 2.4 GeV/c average luminosities of up to $10^{32}$ cm$^{-2}$ s$^{-1}$ are feasible.

## CONCLUSION

The main restriction for the high-resolution mode is intra-beam scattering. To reach the specified momentum spread, the beam has to be heated transversely. This is also required to get a sufficient beam-target overlap. In addition, simulation codes have to be improved to describe the dynamics of tail particles, especially with respect to the beam-target interaction. Beam losses are of major concern for the high-luminosity mode. Hadronic interaction, single Coulomb scattering, energy loss straggling and Touschek effect due to single intra-beam scattering are main causes of the beam loss. At lower momenta, the beam losses are too large compared to the antiproton production rate. An optimized beam cooling scenario and a factor of two larger longitudinal ring acceptance is required to reach average luminosities on the order of $10^{32}$ cm$^{-2}$ s$^{-1}$.

## ACKNOWLEDGEMENTS

This work is supported by INTAS grant No. 03-54-5584 (INTAS Research Project, Advanced Beam Dynamics for Storage Rings).

## REFERENCES

1. An International Accelerator Facility for Beams of Ions and Antiprotons, Conceptual Design Report, GSI Darmstadt, November 2001, see http://www.gsi.de/GSI-Future/cdr/.
2. FAIR Project (subproject HESR), Technical Report, to be published.
3. Strong Interaction Studies with Antiprotons, Letter-of-Intent, PANDA collaboration, January 2004, see http://www.gsi.de/documents/DOC-2004-Jan-1151.pdf.
4. A Study of Spin-dependent Interactions with Antiprotons, Letter-of Intent, ASSIA collaboration, January 2004, see http://www.gsi.de/documents/DOC-2004-Jan-152-1.pdf.
5. Antiproton-Proton Scattering Experiments with Polarization, Letter-of-Intent, PAX collaboration, January 2004, see http://www.gsi.de/documents/DOC-2004-Jan-1251.pdf.

6. F. Rathmann et al., Phys. Rev. Lett. 93, 224801 (2004).
7. A. Lehrach et al., Polarized beams in the high-energy storage ring of the future GSI project, Proc. of the 6th International Spin Physics Symposium, Trieste (2004).
8. Y. Senichev et al., Lattice Design Study for HESR, Proc. of the European Accelerator Conference, Lucerne, 653 (2004).
9. FAIR Project (subproject CR, RESR), Technical Report, to be published.
10. V.V. Parkhomchuk et al., An Electron Cooling System for the Proposed HESR Antiproton Storage Ring, Proc. of the European Accelerator Conference, Lucerne, 1869 (2004).
11. D. Reistad, HESR Electron Cooling System Proposal, to be published in Proc. of International Workshop on Beam Cooling and Related Topics, Galena IL, USA (2005).
12. Ö. Nordhage et al., A high-density pellet target for antiproton physics with PANDA, Proc. of 6th International Conference on Nuclear Physics at Storage Rings, Bonn (2005).
13. A. Dolinskii et al., Simulation Results on Cooling Times and Equilibrium Parameters for Antiproton Beams in the HESR, Proc. of the European Accelerator Conference, Lucerne, 1969 (2004).
14. H. Stockhorst et al., Cooling Scenario for the HESR Complex, to be published in Proc. of International Workshop on Beam Cooling and Related Topics, Galena IL, USA (2005).
15. O. Boine-Frankenheim et al., Cooling equilibrium and beam loss with internal targets in the High-Energy Storage Ring (HESR), to be published.
16. V.V. Parkhomchuk, Nucl. Inst. Meth. A 441, 9 (2000).
17. A. H. Sørensen, CERN Accelerator School (edited by S. Turner) CERN 87-10, 135 (1987).
18. F. Hinterberger, T. Mayer-Kuckuk and D. Prasuhn, Nucl. Instr. Meth. A 275, 239 (1989).
19. F. Hinterberger, D. Prasuhn, Nucl. Instr. Meth. A 279, 413 (1989).
20. INTAS Research Project, Advanced Beam Dynamics for Storage Rings, INTAS Ref. No. 03-54-5584.
21. D. Reistad et al., Measurements of electron cooling and electron heating at CELSIUS, Workshop on Beam Cooling, Montreux (1993), CERN 94-3, p. 183; Yu. Korotaev et al., Intensive Ion Beam in Storage Rings with Electron Cooling, Proc. of IX Russian Particle Accelerator Conference, Dubna, Russia (2004).
22. A.U. Luccio et al., Effect of the electron beam of the cooler for HESR on the machine optics, Proc. of 3rd ICFA Advanced Beam Dynamics Workshop, AIP Conf. Proc. 773, 434 (2005).
23. P. Zenkevich, A. Dolinskii and I. Hofmann, Nucl. Inst. Meth. A 532, 454 (2004).
24. D. Groom, Energy loss in matter by heavy particle, Particle Data Group note PDG-93-06 (1993); H. Bichsel, D.E. Groom, and S.R. Klein, Passage of particles through matter, section 27 in S. Eidelman et al., Review of particle physics, Phys. Lett. B 592, 1 (2004).